\documentclass[aip,graphicx,reprint]{revtex4-1}

\usepackage[sort&compress]{natbib}
\usepackage{amssymb}
\usepackage{booktabs} 
\usepackage[T1]{fontenc}
\usepackage[ngerman,english]{babel}
\usepackage{amsfonts}
\usepackage{amsmath}
\usepackage{array}
\usepackage{epsf}
\usepackage{epsfig}
\usepackage{float}
\usepackage[version=3]{mhchem}
\usepackage{textcomp}
\usepackage{graphics}
\usepackage{graphicx}
\usepackage{epstopdf}
\usepackage{placeins}
\usepackage{footnote}
\usepackage{bbm}

\usepackage{color}

\begin{document}

\title{Multi-reference approach to the calculation of photoelectron spectra including spin-orbit coupling} 

\author{Gilbert Grell}
\author{Sergey I. Bokarev}
\email{sergey.bokarev@uni-rostock.de}
\affiliation{
	Institut f\"{u}r Physik, Universit\"{a}t Rostock, 
	Universit\"{a}tsplatz 3, 18055 Rostock, Germany}
\author{Bernd Winter}
\author{Robert Seidel}
\affiliation{Joint Laboratory for Ultrafast Dynamics in Solutions and at Interfaces (JULiq), Institute of Methods for Material Development, Helmholtz-Zentrum Berlin f\"{u}r Materialien und Energie, Albert-Einstein-Strasse 15, 12489 Berlin, Germany}
\author{Emad F. Aziz}
\affiliation{Joint Laboratory for Ultrafast Dynamics in Solutions and at Interfaces (JULiq), Institute of Methods for Material Development, Helmholtz-Zentrum Berlin f\"{u}r Materialien und Energie, Albert-Einstein-Strasse 15, 12489 Berlin, Germany}
\affiliation{Department of Physics, Freie Universit\"{a}t zu Berlin, Arnimalle 14, 14159 Berlin, Germany}
\author{Saadullah G. Aziz}
\affiliation{Chemistry Department, Faculty of Science, King Abdulaziz University, 21589 Jeddah, Saudi Arabia}
\author{Oliver K\"{u}hn}
\affiliation{
	Institut f\"{u}r Physik, Universit\"{a}t Rostock, 
	Universit\"{a}tsplatz 3, 18055 Rostock, Germany}

\date{\today}

\begin{abstract}
X-ray photoelectron spectra provide a wealth of information on the electronic structure. The extraction of molecular details requires adequate theoretical methods, which in case of transition metal complexes has to account for effects due to the multi-configurational and spin-mixed nature of the many-electron wave function.
Here, the Restricted Active Space Self-Consistent Field method including spin-orbit coupling is used to cope with this challenge and to calculate  valence and core photoelectron spectra. The intensities are estimated within the frameworks of the Dyson orbital formalism and the  sudden approximation. Thereby, we utilize an efficient computational algorithm that is based on a biorthonormal basis transformation. The approach is applied to the valence photoionization of the gas phase water molecule and to the core ionization spectrum of the $\text{[Fe(H}_2\text{O)}_6\text{]}^{2+}$ complex. The results show good agreement with the experimental data obtained in this work, whereas the sudden approximation demonstrates distinct deviations from experiments.
\end{abstract}


\maketitle 

\section{Introduction}
\label{sec:intro}

Spectroscopy, tracing the changes of the energetic levels upon various physical interactions and in course of dynamical processes, is one of the most powerful analytical tools finding applications in almost every field of natural sciences, medicine, and engineering. Among the various spectroscopic methods, experiments in the X-ray regime are suitable to explore the structure of materials in all aggregation states.~\cite{de_groot_core_2008} The most popular variants of X-ray spectroscopies comprise first order absorption and photoelectron spectroscopy as well as second order fluorescence and Auger spectroscopy.~\cite{de_groot_core_2008, milne_recent_2014}

Upon X-ray irradiation, a core hole is created either resonantly or off-resonantly, i.e.\ an electron is excited to a bound or continuum state. Since the binding energies of core electrons are significantly different for various elements and core orbitals have a very localized probability density, X-ray spectroscopy can be used as an element specific local probe of the electronic structure of an atom in its environment. This in particular distinguishes spectroscopy in the  X-ray and UV/VIS regime, since the latter gives insight into transitions between delocalized molecular orbitals (MO).~\cite{stolow04_1719}

Remarkably, photoelectron spectroscopy was found to be sensitive to specific solute-solvent interactions of transition metal (TM) complexes in solutions \cite{winter_liquid_2009,yepes_photoemission_2014, seidel_valence_2011, moens_energy_2010,thurmer_ultrafast_2011} which are essential for understanding processes in catalysis, biochemistry, and material sciences.~\cite{da_silva_biological_1991, wohrle_metal_2003,guo_x-ray_2002,bergmann_isotope_2007,lange_x-ray_2012,ball_water_2008,goodsell_machinery_1993} However, due to the complex electronic structure and notable (especially for core ionization) spin-orbit coupling (SOC) of TM compounds, photoelectron spectra (PES) are rich of features, and an unambiguous assignment is difficult without the aid from theoretical calculations.~\cite{de_groot_core_2008}

On the theory side, a number of methods  has been proposed for the assignment of PES. The simplest ones are Hartree-Fock or Kohn-Sham density functional theory.~\cite{Koerzdoerfer_2009,Koerzdoerfer_2010} Here, MO energies are attributed to the PES transition energies via Koopmans' theorem \cite{Almbladh_exact_1985} and intensities are not analyzed. In some cases, this may provide already a good interpretation of the observed spectrum.~\cite{seidel_valence_2011,yepes_photoemission_2014} However, such a simple picture is not able to describe more complex effects such as combination transitions. For that purpose, methods based on a Green's function approach have been introduced, see e.g.\ Ref.~\citenum{cederbaum_theoretical_1977}. For instance, the algebraic diagrammatic construction formulation \cite{schirmer_new_1983} enjoys particular popularity.~\cite{pernpointner_effect_2005,kryzhevoi_core_2009,santra_x-ray_2009} These methods directly deliver the spectroscopic observables, i.e.\ transition energies and intensities, as poles and residues of the Green's function,~\cite{cederbaum_theoretical_1977} respectively, avoiding the calculation of the stationary wave functions. The relativistic treatment necessary for TM compounds has been implemented  within a four-component formalism \cite{pernpointner_jahnteller_2009} which is quite computationally demanding. 

In principle, any quantum chemical method capable of describing excited electronic states can be used to obtain the PES peak positions of main and combination transitions as energy differences between the $N$-electron initial and $N-1$-electron final states. Methods which have been previously used in PES calculations include Configuration Interaction (CI),~\cite{Sankari_2006,Lisini_1988,Decleva_2009,Arneberg_1982,Kivimaki_2008} Time-Dependent Density Functional Theory (TDDFT),~\cite{di_valentin_gas-phase_2014,mignolet_probing_2013} Equation-of-motion Coupled Cluster (EOM-CCSD),~\cite{melania_oana_dyson_2007,oana_cross_2009} and multiconfigurational  methods based on Complete or Restricted Active Space Self-Consistent Field (CASSCF/RASSCF) wave function.~\cite{Ponzi_2014,josefsson_collective_2013,klooster_calculation_2012} The latter group is of particular importance for TM compounds, since they are known to sometimes have  wave functions which, even in the ground state, cannot be represented by a single configuration.~\cite{roos_multiconfigurational_1996} Alternatively, the semi-empirical valence bond Ligand-Field Multiplet (LFM) technique \cite{Thole_LFM,de_groot_core_2008} is widely used for X-ray photoelectron spectroscopy of TM compounds. In all these methods, the SOC is either not included or treated within the multi-configurational Dirac-Fock method in $jj$-coupling limit \cite{Jankala_2007} or within semi-empirically parametrized (LFM \cite{Thole_LFM,de_groot_core_2008}) and ab initio (RASSCF \cite{josefsson_collective_2013,klooster_calculation_2012}) $LS$-coupling limit.

The intensities in LFM and most of CI studies are estimated in the sudden approximation (SA) \cite{Manne_sudden, aberg_theory_1967} neglecting the kinetic energy dependence of the transition strength and approximating it in a form of a wave function overlap, see Sec.~\ref{sec:theory}. In principle, the SA should be valid for high kinetic energies of the outgoing electron.  However, for certain applications the SA has been shown to be unable to make reliable prediction of intensities.~\cite{Ponzi_2014,Arneberg_1982}

For TDDFT,~\cite{di_valentin_gas-phase_2014,mignolet_probing_2013} EOM-CCSD,~\cite{melania_oana_dyson_2007,oana_cross_2009} CI,~\cite{Decleva_2009} and CASSCF \cite{Ponzi_2014} based methods the Dyson Orbital (DO) formalism~\cite{Pickup_1977} has been applied. Although computationally more demanding, it gives a rather accurate description of intensities. Further, it provides a compact representation of the PES matrix elements by virtue of reducing the $N$-particle to a one-particle integration, see Sec.~\ref{sec:theory}. Finally, DOs can be rigorously employed for the determination of angular-resolved PES.~\cite{reid_p_2003}

The present work sets the focus onto formulating and testing a simulation protocol for the incorporation of SOC and multi-reference effects into the DO formalism for accurate description of PES intensities of TM complexes.  Thereby,  we employ the RASSCF multi-reference approach \cite{roos_complete_1980, olsen_determinant_1988, malmqvist_restricted_1990}  and include SOC within state interaction (RASSI) method \cite{malmqvist_restricted_2002} in the atomic mean field integral  approximation.~\cite{schimmelpfennig_amfi_1996} The paper is organized as follows. In Sec.~\ref{sec:theory} we give a general overview of the theoretical background and of the developed methodology. Computational details are provided in Sec.~\ref{sec:compdet} and experimental details in Sec.~\ref{sec:exp}. Section~\ref{sec:res} discusses application of the  described protocol  to two model systems, i.e. gas phase water (valence PES) and the $[\text{Fe}(\text{H}_2\text{O})_6]^{2+}$ complex (core PES), which corresponds to solvated iron(II) ions in water. In Sec. \ref{sec:concl} we conclude that the proposed approach allows the description of PES on the same footing as processes involving photons, i.e.\ X-ray absorption and resonant inelastic scattering, reported recently.~\cite{josefsson_ab_2012,wernet_dissecting_2012,suljoti_direct_2013,bokarev_state-dependent_2013,atak_nature_2013,engel_chemical_2014,pinjari_restricted_2014,Wernet_2015}

\section{Theory}
\label{sec:theory}
We consider neutral or charged molecules with $N$ electrons, which are initially in their ground state $|\Psi_{I}^N\rangle$. In the sudden ionization limit,~\cite{Manne_sudden, Pickup_1977} the final state can be written as an antisymmetrized product $|\Psi_F^{N-1}\psi^{\rm el}(\mathbf{k})\rangle$ of the continuum state of the photoelectron $\psi^{\rm el}(\mathbf{k})$ and the wave function of the $N-1$-electron remainder $\Psi_{F}^{N-1}$.

Assuming that all photoelectrons with a certain kinetic energy ($\mathcal{E}_{k}=\hbar^2k^2/(2m_{\rm e})$) are detected regardless of their outgoing direction and spin, their number per unit time is proportional to the transition rate, including all possible initial and final states with energies $\mathcal{E}_I$ and $\mathcal{E}_F$ and integrated over all directions, $d\Omega_k$, of the outgoing electron. In the long wavelength approximation, it reads:
\begin{equation}
\label{expPES}
\begin{split}
\sigma(\mathcal{E}_{k})&\propto\frac{2\pi}{\hbar}\sum_I{f^{\rm B}(\mathcal{E}_I)}\sum_F \Lambda (\mathcal{E}_F+\mathcal{E}_{k}-\mathcal{E}_I-\hbar\omega)\\
&\times\int{d\Omega_k}\left|\left\langle \Psi_F^{N-1}\psi^{\rm el}(\mathbf{k})\left|\vec{\varepsilon}\cdot\hat{\mathbf{d}}\right|\Psi_I^N\right\rangle\right|^2,
\end{split}
\end{equation}
where $\hat{\mathbf{d}}$ is the dipole operator, $\vec{\varepsilon}$ is the polarization of the incoming photon, the lineshape function $\Lambda(\mathcal{E})$ accounts for the finite width of the excitation pulse, inhomogeneous, and other broadening effects. The thermal population of the initial states enters Eq.~\eqref{expPES} via the Boltzmann factor $f^{\rm B}(\mathcal{E}_I)$.

In the following we will specify the many-body states $\Psi^N_I$ and $\Psi^{N-1}_F$ in terms of Slater determinants (SD), $\Theta_i$, composed of single particle MOs, $\varphi_i^k$. Let us consider the PES matrix elements in Eq.~\eqref{expPES}  for  two  SDs giving a contribution to the initial state $\Psi^N_I$, i.e.\ $\Theta_j^N$, and the final state $\Psi^{N-1}_F$, i.e.\ $\Theta_i^{N-1}$. Omitting for convenience the polarization and $\mathbf{k}$ dependence and applying the strong orthogonality condition between the free electron function and initial molecular orbitals ($\left\langle\psi^{\rm el}|\varphi_j^k\right\rangle=0$, see, e.g., Ref.~\citenum{Ritchie_orthogonal}) one can write
\begin{equation}
\label{PES1}
D_{ij}=\left\langle\Theta_{i}^{N-1}\psi^{\rm el}\left | \hat{\mathbf{d}}\right |\Theta_{j}^N\right\rangle= \left\langle\psi^{\rm el}\left | \hat{\mathbf{d}}\right |\Phi_{ij}\right\rangle \, ,
\end{equation}
where we introduced the DO
\begin{eqnarray}\label{SD_DO}
\Phi_{ij}^{\rm SD}&=&\sum_{\mathcal{P}\in S_N}{(-1)^p}\Bigl\langle \varphi_i^{1}\Bigl|\varphi_j^{\mathcal{P}(1)}\Bigr\rangle\cdots\left\langle\varphi_i^{N-1}\Bigr|\varphi_j^{\mathcal{P}(N-1)}\right\rangle \nonumber\\
&\times & \varphi_j^{\mathcal{P}(N)}\, .
\end{eqnarray}
Here, $\mathcal{P}\in S_N$ denotes  all possible permutations of $N$ orbital indices with parity $p$ coming from the structure of the SDs. Thus, the expression for the PES matrix element simplifies to a one-particle integration (electron coordinates ${\bf x}_k$) because the $N-1$-dimensional integration is comprised into the  DO with the normalization factor $\sqrt{N}$
\begin{equation}\label{dys}
\begin{split}
\Phi_{ij}^{\rm SD}(\mathbf{x}_N)&=\sqrt{N}\int{\left(\Theta_i^{N-1}(\mathbf{x}_1,\cdots,\mathbf{x}_{N-1})\right)^{\ast}}\\
&\times\Theta_j^{N}(\mathbf{x}_1,\cdots,\mathbf{x}_{N})d\mathbf{x}_1\cdots d\mathbf{x}_{N-1}\, .
\end{split}
\end{equation}
Note that in general the one-electron orbitals of the initial $N$-electron and final $N-1$-electron bound states are not orthonormal due to electronic relaxation upon electron removal. The DO can be considered as an analogue of the reduced one-electron transition density. Its direct correspondence is the wave function of the photoelectron before the ionization. Further, the DO is not normalized and the PES intensity is proportional to its norm. The convenience of the DO formulation of the PES matrix elements comes with the fact that the DO contains all system- and method-specific information. Therefore, one can vary the approximations for $\psi_{\rm el}$, without changing the DO.

In practice, the $N!$-fold summation in Eq.~\eqref{SD_DO} can be circumvented by calculating $N$ determinants of $N-1 \times N-1$ dimensional overlap matrices, similar to Ref.~\citenum{aberg_theory_1967}. Thus the computational effort reduces to $\mathcal{O}(N^4)$.  A more efficient scheme can be obtained upon transformation of the  nonorthonormal MO basis sets and CI coefficients of the initial and final states $\Psi_I^N$  and $\Psi_F^{N-1}$ to biorthonormal sets.~\cite{malmqvist_restricted_2002, malmqvist_calculation_1986} This  strongly simplifies the DO expression in Eq.~\eqref{SD_DO}, which reduces to a single term. To show this we consider the case where the removed electron was in orbital $\varphi_j^{\tau}$ in the initial state $\Theta_j^N$. If  this orbital is not contained in the final state $\Theta_i^{N-1}$, then only one permutation in Eq.~\eqref{SD_DO} leads to nonzero results. That is the permutation $\mathcal{P}$, which shifts the electron in $\varphi_j^{\tau}$ into the $N$th half space of the initial state which does not take part in the integration. The parity of the permutation $\mathcal{P}$ is $p=N-\tau$. Thus, we arrive at a simple expression for the DO in the biorthonormal basis
\begin{equation}
\label{DObiofin}
\Phi_{ij}^{\rm SD}=(-1)^{N-\tau}\varphi_j^{\tau} \, .
\end{equation}

The most important feature of the representation in the biorthonormal basis is that the DO for a pair of determinants corresponds to a spin-orbital of the initial SD taken with the appropriate sign. Thus the computational effort does not depend on the number of electrons, which makes this approach very efficient. 

In the single SD picture, the DO depends on the coordinates of only one electron. Thus, no combination transitions, where one electron is ejected and others are simultaneously excited, are taken into account. However, in a multi-reference description of the DO, these effects are included. To obtain wave functions, which account for the multi-reference character of transition metal complexes, as well as SOC effects, we follow a two-step strategy. First, the spin-free RASSCF wave functions having  ground state spin $S$ as well as $S\pm 1$ are calculated. They have the form of a linear combination of different SDs $\Theta_i$, weighted with CI coefficients $C_{Ki}$ (omitting the superscript for the number of electrons in the following)
\begin{equation}
\label{CIwfSD}
\Psi^{\rm CI}_{K=F,I}=\sum_{i=1}^{N_{\rm CI}}{C_{Ki}\Theta_i}.
\end{equation}

Assuming that the initial state can be described by a single SD ($\Theta_{0}$), the DO within the CI approach, omitting for convenience $I$ and $F$ indices, can be written in the form  
\begin{equation}
\label{corrDO}
\begin{split}
\Phi^{\rm CI}_{FI}&=\sum_i{\underbrace{C^{\ast}_{i} C_{0}\left\langle\Theta_{i}\bigr|\Theta_{0}\right\rangle}_{\rm 1h}}+\sum_{ija}{\underbrace{\left(C^a_{ij}\right)^{\ast}C^a_{j}\left\langle\Theta_{ij}^{a}\bigr|\Theta_{j}^{a}\right\rangle}_{\rm 2h1p}}\\
&+\sum_{ijkab}{\underbrace{\left(C^{ab}_{ijk}\right)^{\ast}C^{ab}_{jk}\left\langle\Theta_{ijk}^{ab}\bigr|\Theta_{jk}^{ab}\right\rangle}_{\rm 3h2p}}+\ldots \,.
\end{split}
\end{equation}
Here, the brakets denote the $N-1$-dimensional integration, $i,j,k,\ldots$ are indices of the occupied orbitals from which the electrons are removed, $a,b,\ldots$ correspond to the unoccupied orbitals into which the electrons are moved. The first three sums represent the main 1h (single hole) and combination 2h1p (two holes, single particle) and 3h2p  (three holes, two particles) transitions, respectively.

The concept of active space within RASSCF as applied to Eq.~\eqref{corrDO} allows to flexibly vary the highest order of correlation terms, in principle up to full CI within the given orbital subspace. However, this subspace needs to be large enough to include all relevant ionization channels. Interestingly, when employing the biorthonormal basis set transformation, orbital relaxation effects, which can cause satellites due to non-orthogonality of initial and final orbitals \cite{Lisini_1988}, are shifted completely to correlation effects, i.e.\ the CI expansion. 

At the second step, these wave functions are coupled with the  SOC operator in atomic mean-field integral approximation \cite{schimmelpfennig_amfi_1996} to generate spin-orbit wave functions within the RASSI  approach.~\cite{malmqvist_restricted_2002, malmqvist_calculation_1986} These spin-orbit wave functions are expanded in terms of the spin-free states $\Psi_{Kn,M_\sigma, \sigma}^{CI}$, where $K=F,I$, (Eq.~ \eqref{CIwfSD}) with spin $\sigma$ ($N_\sigma$ states in total), magnetic quantum numbers $M_{\sigma}$ and SOC coefficients $\xi_{Kn,M_{\sigma},\sigma}^{}$:
\begin{equation}
\label{SOCWF}
\Psi^{\rm SO}_{K=F,I}=\sum_{\sigma=S-1}^{S+1}{\sum_{n=1}^{N_{\sigma}}}\sum_{M_\sigma=-\sigma}^{\sigma}\xi_{Kn,M_{\sigma},\sigma}^{}\Psi_{Kn,M_{\sigma},\sigma}^{\rm CI}
\end{equation}
The particular choice of spin manifolds is dictated by the SOC selection rules $\Delta S=0,\pm1$.

Thus, the multi-configurational DO including SOC for the transition from initial state $I$ to final state $F$  can be expanded in terms of single determinant DO $\Phi_{ij}^{\rm SD}$ (Eq.~\eqref{SD_DO})
\begin{equation}
\label{soflatDO}
\Phi^{\rm SO}_{FI}=\sum_{k=1}^{N_{F}^{\rm SO}}\sum_{l=1}^{N_{I}^{\rm SO}}\xi_{Fk}^{\ast}\xi_{Il}\sum_{i=1}^{N_F^{\rm SD}}\sum_{j=1}^{N_I^{\rm SD}}C_{Fi}^{\ast}C_{Ij}\Phi_{ij}^{\rm SD}.
\end{equation}
Here, the nested sums over spin, magnetic quantum number and spin-free states are replaced by one sum over all different spin-orbit states with the total number of final and initial SOC states $N_{F}^{\rm SO}$ and $N_{I}^{\rm SO}$, respectively. The above Eq.~\eqref{soflatDO} can be used to calculate the PES matrix elements $D_{FI}^{\rm SO}=\bigl\langle \psi^{el}|\hat{\mathbf{d}} | \Phi^{\rm SO}_{FI} \bigr\rangle$, which is the main working expression used in this work.

\begin{figure}[t]
\includegraphics{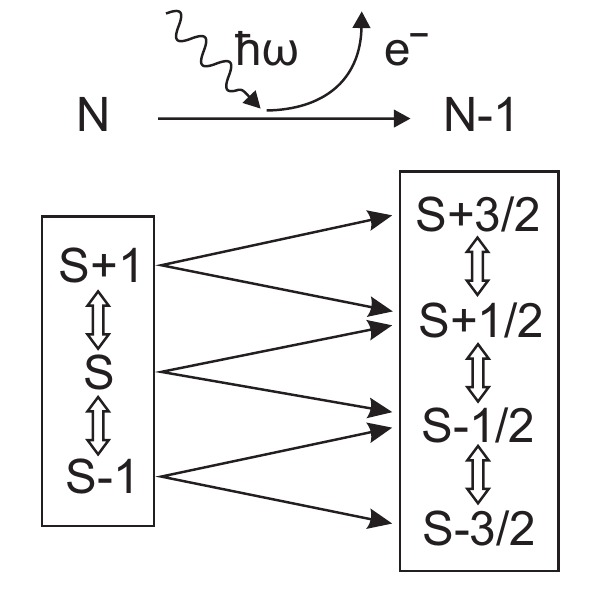}
\caption{\label{rassiSch}
Schematic of the directly spin-coupled manifolds of states (the double-sided arrows, $\Delta S=0, \pm 1$) of the $N$ and $N-1$ electron systems. Depending on the spin of the outgoing electron, the spin of final $N-1$ electron states changes by $\Delta S=\pm 1/2$ (single-sided arrows).}  
\end{figure}

Since an electron with either $\alpha$ or $\beta$ spin is removed from the initial spin-free states with spins $S$ and $S\pm1$, the final states with $\Delta S=\pm 1/2$ need to be considered. That is why in general seven spin manifolds of the unionized molecule with $N$ electrons and its $N-1$-electron ion need to be taken into account as depicted in Fig.~\ref{rassiSch}.

Frequently, the SA is used for the calculation of the transition dipole matrix element in Eq.~\eqref{expPES}.~\cite{Manne_sudden, aberg_theory_1967} Here, the matrix element  is approximated as an overlap integral neglecting the $\mathbf{k}$-dependence of the transition strength
\begin{equation}
D_{FI}^{\rm SA}=\left|\left\langle \Psi_F^{N-1}\psi^{\rm el}(\mathbf{k})\left|\hat{\mathbf{d}}\right|\Psi_I^N\right\rangle\right|^2 \approx\left|\left\langle \Psi_F^{N-1}\Bigr|\hat{a}\Psi_I^N\right\rangle\right|^2 \, ,
\end{equation}
where the operator $\hat{a}$ annihilates the electron from the occupied MO. Expressed in terms of DOs in the SA approach, the one-electron integration for $\langle\psi^{\rm el} | \hat{\mathbf{d}} |\Phi_{FI}\rangle$ is omitted and the intensity is approximated by the DO norm, $|\Phi_{FI}|^2$, only. Most of the published papers use the SA for the prediction of the intensities of the combination transitions relative to the intensities of main lines. However, the accuracy of the predicted intensity ratios between different main transitions occuring at high excitation energies has been questioned in Refs.~\citenum{Palma_1980,Arneberg_1982}.

\section{Computational details}
\label{sec:compdet}
The spectra were calculated for two test systems, i.e.\ gas phase water and the $[\text{Fe}(\text{H}_2\text{O})_6]^{2+}$ complex mimicking the $\text{Fe}^{2+}$ ion within its first solvation shell in aqueous solution. Thus the following processes were studied
\[
\text{H}_2\text{O}^{0}\longrightarrow\text{H}_2\text{O}^{+}+e^-\\
\]
\[
[\text{Fe}(\text{H}_2\text{O})_6]^{2+}\longrightarrow\text[\text{Fe}(\text{H}_2\text{O})_6]^{3+}+e^-\,.
\]
The geometry of water was optimized at the density functional theory level with the B3LYP functional \cite{becke_density-functional_1993, lee_development_1988} together with the 6-311G(d) basis set.~\cite{mclean_contracted_1980,krishnan_self-consistent_1980}
For the $[\text{Fe}(\text{H}_2\text{O})_6]^{2+}$ complex the geometry was first optimized using the MP2 method with the cc-pVTZ basis set.~\cite{dunning_gaussian_1989, woon_gaussian_1994, balabanov_systematically_2005} Then the $\text{Fe}-\text{O}$ bond lengths were set to the  CASPT2$(10e^-,12\text{MO})$/ANO-RCC values obtained in Ref.~\citenum{pierloot_relative_2006}.
All geometry optimizations were done with Gaussian 09 program package.~\cite{g09}

\begin{figure*}[t]
	\includegraphics{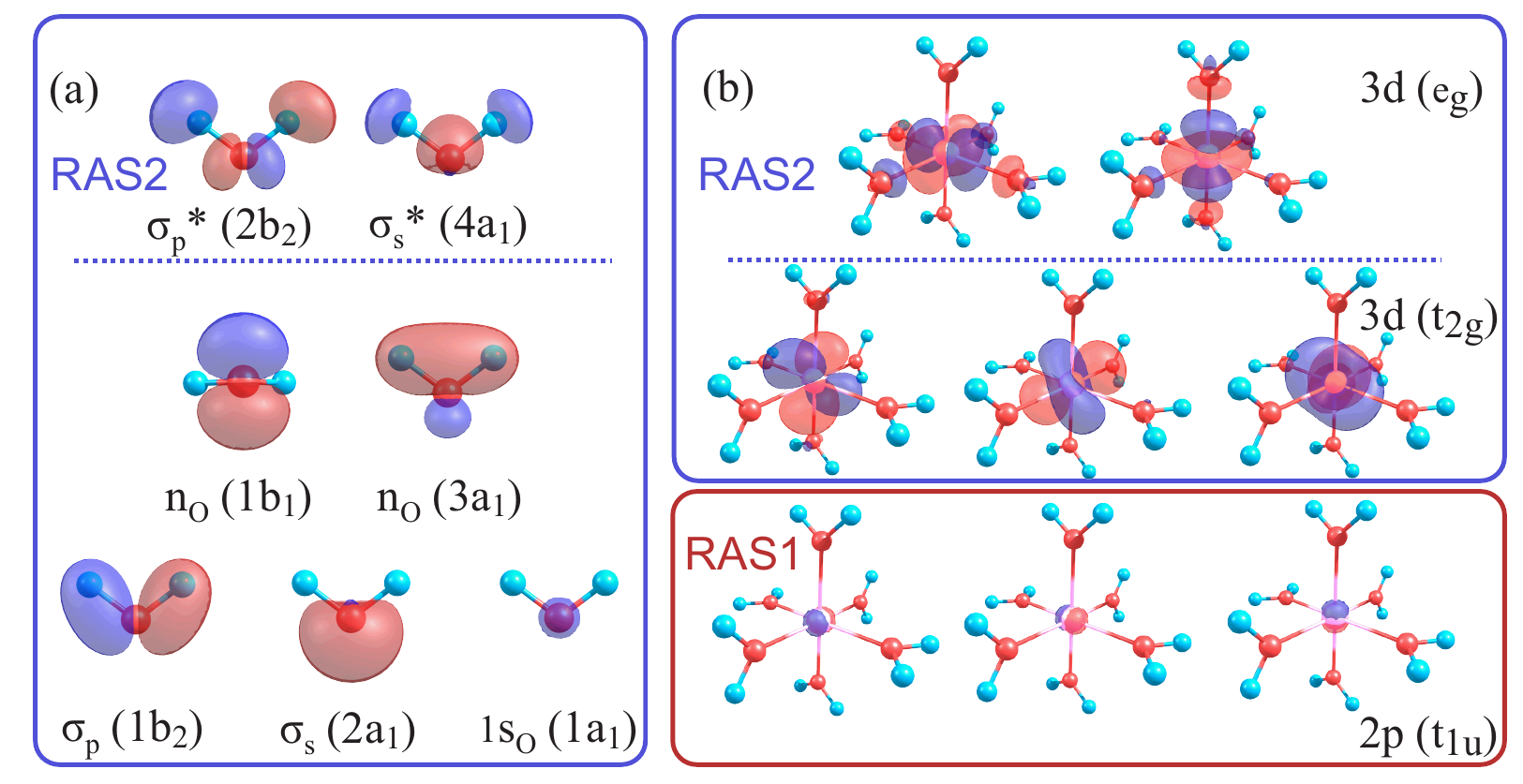}
	\caption{\label{AS}
		Active spaces used for RASSCF calculations of (a) water and (b) $[\text{Fe}(\text{H}_2\text{O})_6]^{2+}$.}  
\end{figure*}

For state-averaged RASSCF calculations of water, the full valence RAS2 active space additionally including the 1s core orbital of oxygen was chosen (Fig.~\ref{AS}). For the iron complex, the active space included the 3d orbitals of iron in the RAS2 space and allowed single electron excitations from the iron 2p core orbitals in the RAS1 space. This results in the singly core-excited or ionized states, having full flexibility in the valence 3d manifold. This active space corresponds to that used for the study of X-ray absorption and resonant inelastic scattering spectra  \cite{bokarev_state-dependent_2013} providing interpretation on the same footing. The wave function was first optimized for the ground state and then all orbitals apart from the active ones were kept frozen. 

The (8s4p3d)/[3s2p1d] ANO-RCC basis on hydrogen and (14s9p4d3f)/[5s4p3d2f] basis on oxygen \cite{roos_new_2005} were employed for water, which corresponds to quadruple-zeta quality. For  $\text{Fe}^{2+}$ the contractions (21s15p10d6f)/[6s5p3d2f], (14s9p4d3f)/[4s3p2d1f], and (8s4p)/[2s1p] were used for iron, oxygen and hydrogen, respectively, corresponding to the triple-zeta level.

To account for dynamic correlation, the second order perturbation theory correction (RASPT2) was calculated \cite{malmqvist_restricted_2008, andersson_second-order_1990} for the case of water. To avoid intruder state singularities the imaginary level shift of 0.4 Hartree was applied. Because of the much larger number of transitions only RASSCF energies were calculated for the iron complex.

To incorporate scalar relativistic effects, the Douglas-Kroll-Hess transformation \cite{douglas_Quantum_1974,hess_relativistic_1986} up to second order was utilized.
For the case of water, SOC was not taken into account and only transitions from the singlet ground state to the 490 doublet singly ionized final states were considered. This corresponds to one out of six possible branches in Fig.~\ref{rassiSch}. In case of the iron complex, quintet ($S=2$) and triplet ($S=1$) initial states as well as sextet ($S=5/2$), quartet ($S=3/2$), and doublet ($S=1/2$) final states were taken into account. Septet ($S=3$) and octet ($S=7/2$) states are not possible with the active space chosen here. However, it was shown that they play a very minor role for X-ray absorption spectrum  \cite{bokarev_state-dependent_2013} and can be neglected. Thus, only five branches out of the seven in Fig.~\ref{rassiSch} were considered. In total, for $[\text{Fe}(\text{H}_2\text{O})_6]^{2+}$ one initial ground and 1260 core-excited SOC states were included. Thermal population of the low lying initial states was neglected. 

The RASSCF/RASPT2/RASSI calculations were done without any symmetry restriction using  a locally modified MOLCAS 8.0 program package.~\cite{aquilante_molcas_2010} 

The transition dipole matrix elements in Eq.~\eqref{PES1} were calculated with the ezDyson 3.0 program \cite{gozem_ezdyson_2015} via numerical integration of the DO with the free electron wave function $\psi^{\rm el}(\mathbf{k})$. Neglecting the interaction between the photoelectron and ionic remainder one may express $\psi^{\rm el}(\mathbf{k})$ as a plane wave expanded in a basis of spherical waves with spherical harmonics $Y_{l,m}(\mathbf{k})$ in $\mathbf{k}$-space as coefficients (see, e.g., Ref.~\citenum{sakurai_modern_2011}). Further, the spherical waves are expanded in position space using spherical Bessel functions $j_l(k \cdot r)$, where $k=|\mathbf{k}|$, $r=|\mathbf{r}|$, and  spherical harmonics $Y_{l,m}(\mathbf{r})$ yielding
\begin{equation}
\label{photoEl}
\psi^{\rm el}(\mathbf{k})=\sum_{l=0}^{\infty}{\sum_{m=-l}^{l}i^l\sqrt{\frac{2}{\pi}}j_l(k\cdot r)Y_{l,m}(\mathbf{r})Y_{l,m}^{\ast}(\mathbf{k})}\, .
\end{equation}
This particular form of the plane wave has the advantage that one can truncate the infinite angular momentum expansion at some $l_{\rm max}$, e.g., according to the dipole selection rule $\Delta l=\pm 1$. The natural choice of maximum angular momentum in the expansion Eq.~\eqref{photoEl} corresponds to the $l_{\rm max}=l_{\rm bas}+1$, where $l_{\rm bas}$ is maximum angular momentum included in the atomic basis set. However, due to lower than spherical symmetry of the molecules under study, sometimes a larger $l_{\rm max}$ needs to be selected, see Ref.~\citenum{melania_oana_dyson_2007} and Sect.~\ref{sec:res}. In case of water, $l_{\rm max}=7$, box with a side length of $10$~$\text{\AA}$ and an equidistant grid of $300\times300\times300$ points for numerical integration of Eq.~\eqref{PES1} ensures convergence of the DO norms with an accuracy of $10^{-5}$. Since the intensity of transition scales quadratically with the norm of the DO, they have been evaluated only, if the norm of the respective DO was larger than  $10^{-3}$. For the iron ion, only contributions from those DOs that have a norm larger than $10^{-2}$ were taken into account. Here $l_{\rm max}=5$, box size  of $16\text{\AA}$ and a grid of $480\times480\times480$ points was used for the numerical integration reproducing the norms of DOs with an accuracy of $10^{-4}$.

%
%
As shown in Sec.~\ref{sec:theory}, the spin-coupled DO comprises contributions from different spin-states. Therefore, the total spin of the DO and outgoing electron is not well defined. The consequence is that the DO consists of both $\alpha$ and $\beta$ spin wave functions
\begin{equation}
\Phi^{\rm SO}_{FI}=\Phi^{\rm SO}_{FI}(\alpha)+\Phi^{\rm SO}_{FI}(\beta)\,,
\end{equation}
which are complex-valued due to the SOC.  Here, we neglect spin-coherence and calculate the squared PES matrix element as
\begin{equation}
\label{PESsoAp}
\Bigl|D_{FI}^{\rm SO}\Bigr|^2\approx\Bigl|D_{FI}^{\rm Re}(\alpha)\Bigr|^2+\Big|D_{FI}^{\rm Im}(\alpha)\Bigr|^2+\Big|D_{FI}^{\rm Re}(\beta)\Bigr|^2+\Big|D_{FI}^{\rm Im}(\beta)\Bigr|^2,
\end{equation}
where Re and Im represent the real and imaginary parts of $D_{FI}$. The broadenings of the spectral lines were fitted to reproduce the experimental data and are discussed in the next section. No nuclear vibrational effects were taken into account.
The PES matrix element was integrated over all possible outgoing directions of the photoelectron and averaged over all possible orientations of the molecules to mimic free tumbling of the solute in a liquid phase.

\section{Experimental details}\label{sec:exp}
The PES were measured from a 15 $\mu$m vacuum liquid jet \cite{winter_photoemission_2006,seidel_photoelectron_2011} at the soft-X-ray U41 PGM undulator beamline of the Berlin synchrotron radiation facility, BESSY II, Berlin. The jet velocity was approximately 100 $\text{m}\cdot{s}^{-1}$, and the jet temperature was $6^{\circ}\text{C}$. At operation conditions, the pressure in the interaction chamber is $\approx 1.5\cdot10^{-4}$ mbar. Electrons were detected normal to both the synchrotron-light polarization vector and the flow of the liquid jet. A 100 $\mu$m diameter orifice that forms the entrance to the hemispherical electron energy-analyzer (Specs Leybold EA10) is typically at approximately 0.5 mm distance from the liquid jet. Because of the small focal size ($12\times23 \ \mu \text{m}^2$) of the incident photon beam at the interaction point with the liquid jet, by moving the jet slightly out of focus the photoelectron spectrum from gas-phase water can be measured. The valence spectrum of the latter was obtained using 180 eV photon energy; the total energy resolution was better than 60 meV. For the aqueous phase $\text{Fe}^{2+}$ 2p core-level spectrum the photon energy was 925 eV, and the energy resolution was approximately 300 meV. The 2M iron aqueous solution (pH $\approx 2.5$) was prepared by adding anhydrous $\text{FeCl}_2$ salt ($98\%$ purity, Sigma Aldrich) to highly demineralized water ($>17 \ \text{MOhm}\cdot \text{cm}^{-1}$) water.
The major species in this pale-lime colored aqueous solution are $[\text{Fe(H}_2\text{O)}_5\text{Cl}]^+$ ($\approx 77\%$) and $[\text{Fe(H}_2\text{O)}_6]^{2+}$ ($\approx 23\%$).~\cite{Heinrich_1990}

\section{Results and Discussion}\label{sec:res}

\subsection{Water}
The PES of water was chosen as a benchmark for the derived protocol for calculation of PES for two reasons. First, the water valence PES is well understood both experimentally and theoretically.~\cite{winter_full_2004,Lisini_1988,Sankari_2006,Palma_1980,Arneberg_1982,Banna_1986}  Second, water is not influenced by strong SOC and has mostly single-configurational character as long as only the lowest valence excitations are regarded.  This substantially simplifies the calculation and analysis of the nature of transitions, since DOs represent mostly single MOs.

\begin{figure}[t]
	\includegraphics{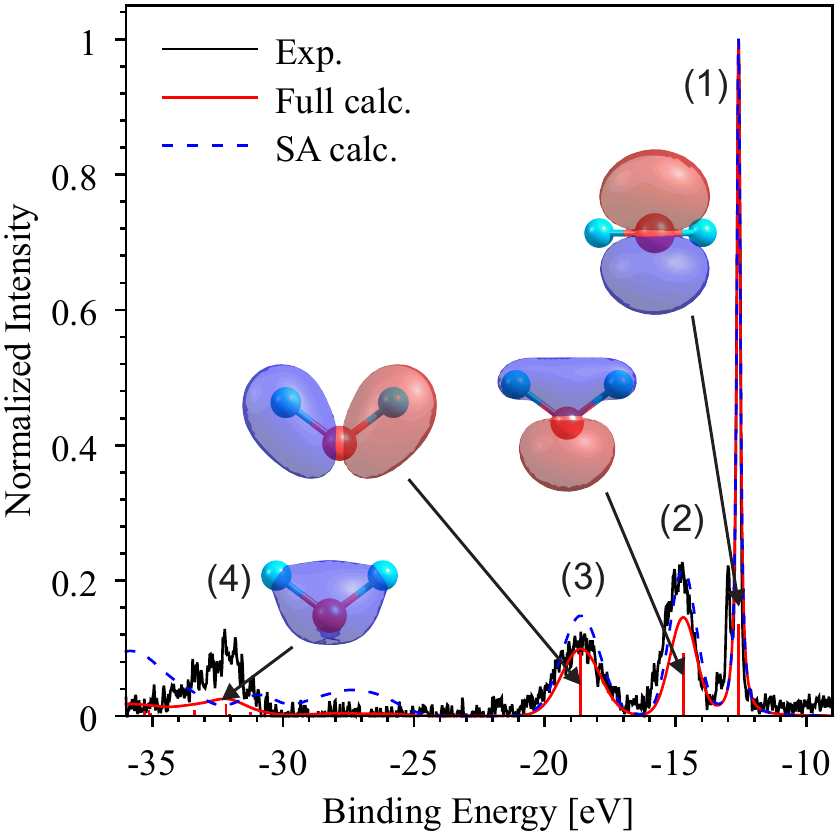}
	\caption{
	Calculated and experimental PES of water in the gas phase for 180~eV excitation energy.	Full calculation corresponds to numerical integration of PES matrix element, SA means sudden approximation.~\cite{Manne_sudden, aberg_theory_1967} Spin-free DOs of the selected transitions are also shown.}
	\label{H2OPES}
\end{figure}

In Fig.~\ref{H2OPES}, the experimental and theoretical PES for the 180~eV energy of the incoming photon are shown.  For convenience, the individual transitions are also given as a stick spectrum.
The normalized intensity is plotted against binding energy $E_b=\mathcal{E}_F-\mathcal{E}_I$, where $\mathcal{E}_F$ and $\mathcal{E}_I$ are the energies of final and initial states, respectively. 
Additionally, the DOs corresponding to selected transitions are presented.
The broadening of each transition was fitted to reproduce the experiments: peak (1) (see Fig. \ref{H2OPES}) Lorentzian lineshape FWHM=0.17~eV, peaks (2) and (3) Gaussian lineshape with FWHM of 1.18~eV and 1.75~eV, respectively. For all other transitions contributing to peak (4), Gaussian profiles with widths 1.5~eV for $E_b<35$~eV  and 3.0~eV for $E_b>35$~eV were applied. The absolute energy shift of -0.78~eV was chosen to align peak (1) with the experiment.

The calculated spectrum is in rather good agreement with the experimental data. There are slight variations of intensities of peaks (2) and (3) relative to (1) and the most notable discrepancies between theory and experiment are observed for peak (4). These deviations in the intensities and lineshapes could be ascribed to the fact that nuclear vibrational effects were not taken into account. However, the relative energetic positions of the peaks are predicted with very high accuracy. To note is the fact that RASPT2 correction is essential here to reproduce the transition energies. This stems from the fact that in RASSCF dynamic correlation is accounted for in an unbalanced way within the active space and may substantially change upon removal of one electron. To correct for this behavior a more complete treatment of correlation, such as RASPT2, is needed.

It can be seen that SA predicts the relative intensities of peaks (1) to (3) with similar quality as  the full calculation employing integration of the DO with the free electron wave function. But, for peak (4) the agreement between SA and experiment is not even qualitative. 

The assignment of peaks fully agrees with that established previously.~\cite{winter_full_2004,Lisini_1988,Sankari_2006,Palma_1980,Arneberg_1982,Banna_1986} Peaks (1)-(3) correspond to ionizations from the lone pairs $n_{\rm O}(1b_1)$ and $n_{\rm O}(3a_1)$ and the $\sigma_{\rm p}(1b_2)$ MO of water, respectively. The fourth peak consists of several transitions of different nature most of them having $\sigma_{\rm s}(2a_1)$ character. This can be ascribed to the multi-configurational character of the wave function and appearance of combination transitions, see Supporting Information. Interestingly, the DOs of $a_1$ symmetry resemble very closely the ground state Hartree-Fock MOs, but in fact they are linear combinations of the $\sigma_{\rm p}(2a_1)$ and $\sigma_{\rm p}(3a_1)$ RASSCF active orbitals (cf.\ Fig.~\ref{AS}).

\begin{figure}[t]
	\includegraphics{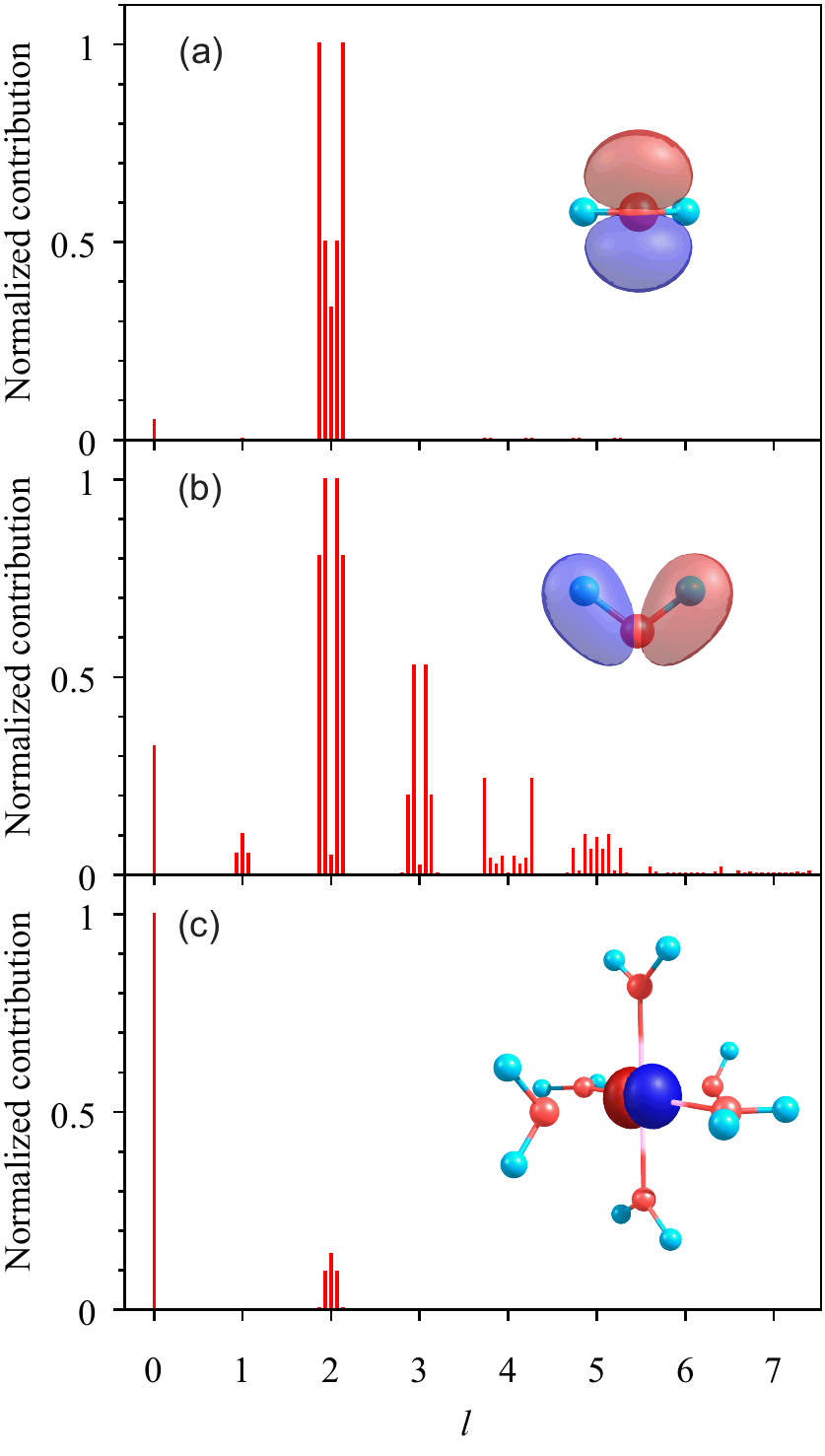}
	\caption{
	Normalized contributions of different $\{l,m\}$ partial waves to the intensity of selected transitions: (a) and (b) transitions (1) and (3) of an isolated water molecule, respectively, see Fig.~\ref{H2OPES}; (c) real part of $\beta$ spin DO contribution of transition (2) of the $[\text{Fe}(\text{H}_2\text{O})_6]^{2+}$ core PES, see Fig.~\ref{FePEScore}. Each group of $2l+1$ sticks corresponds to $m$-components, see Eq.~\eqref{photoEl}. Corresponding DOs are shown as insets using different contour values for visual clarity.
	}
	\label{H2Ocklm}
\end{figure}

The water molecule has rather low point symmetry and convergence of the series in Eq.~\eqref{photoEl} is quite slow.  The contributions of the partial $\{l,m\}$ waves to intensities of peaks (1) and (3) are presented in Fig.~\ref{H2Ocklm}. The DOs of transitions (1) and (2) represent almost pure 2p orbitals of oxygen and in accordance with the dipole selection rules ($\Delta l=\pm 1$) only contributions with $l=\{0,2\}$ notably differ from zero. In contrast, the DOs for transitions (3) and (4) deviate strongly from atomic character, being rather delocalized MOs such that the series in Eq.~\eqref{photoEl}  converges notably slower, approaching zero only for $l\geq6$.  In general, one can conclude that the more the DO is delocalized over the molecule and the less symmetric it is, the more terms have to be included in the photoelectron wave expansion (Eq.~\eqref{photoEl}), see also Ref.~\citenum{melania_oana_dyson_2007}.

\subsection{Fe$^{2+}$ (aq.)}

\begin{figure}[t]
	\includegraphics{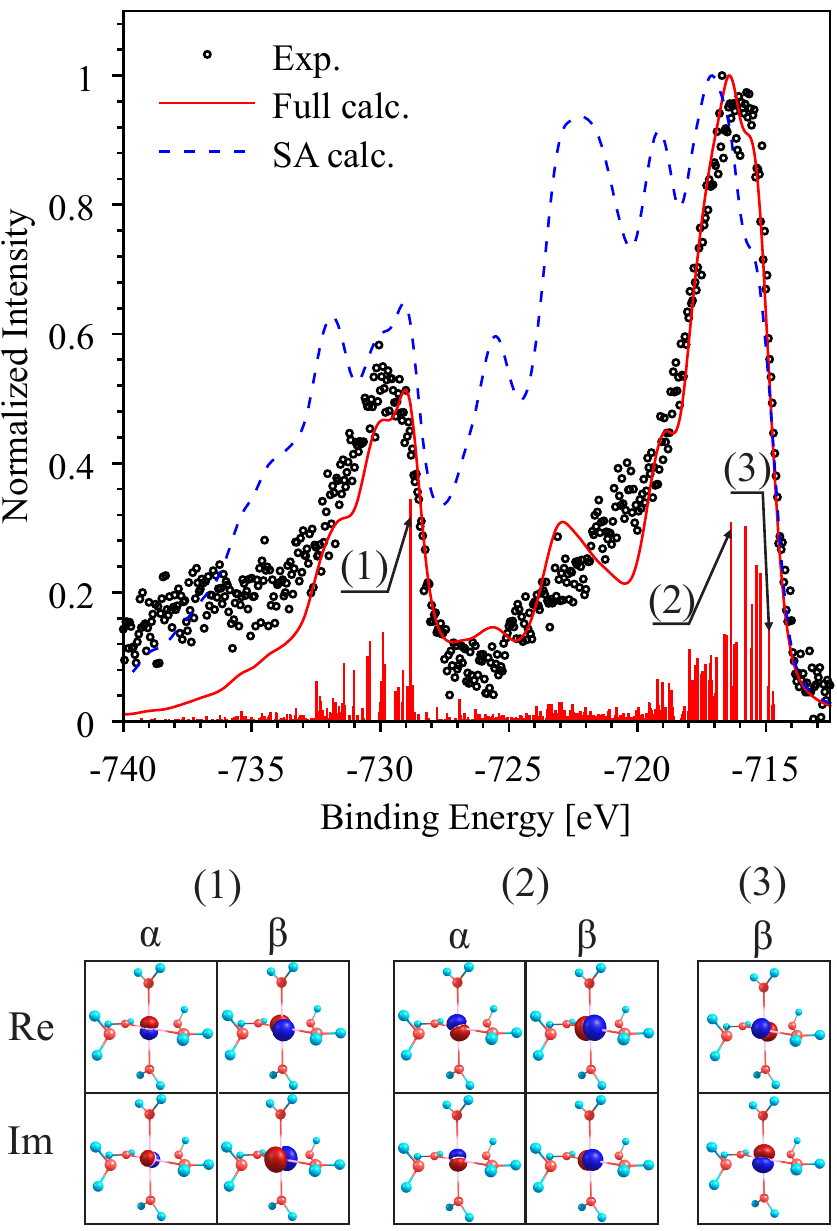}
	\caption{
	(a): Experimental (2M $\text{Fe}\text{Cl}_2$  aqueous solution) and calculated (for $[\text{Fe}(\text{H}_2\text{O})_6]^{2+}$ cluster) core PES for incoming photon energy of 925~eV. Full calculation corresponds to numerical integration of PES matrix element, SA means sudden approximation.~\cite{Manne_sudden, aberg_theory_1967}
	(b): Real and imaginary parts of $\alpha$ and $\beta$ spin contributions to the DOs for selected transitions.}
	\label{FePEScore}
\end{figure}

The calculated L-edge core PES of $[\text{Fe}(\text{H}_2\text{O})_6]^{2+}$ at incoming photon energy of 925~eV is shown in Fig.~\ref{FePEScore}a together with experimental results for a 2M aqueous solution of $\text{Fe}\text{Cl}_2$ at the same energy. The stick spectrum was broadened using the Voigt profile
\begin{equation}
\label{voigt}
V(x)=\int_{-\infty}^{\infty}{G(\sigma,x')L(\gamma,x-x')dx}\,,
\end{equation}
with the Gaussian and Lorentzian lineshape functions $G(\sigma,x)$ and $L(\gamma,x)$.
For the broadening of the $L_3$ peak ($E_b>-727.4$~eV), 0.5~eV and 0.7~eV were used for the Lorentzian and Gaussian FWHM, respectively. For the $L_2$ peak (below $E_b<-727.4$~eV), 0.7~eV for both Lorentzian and Gaussian width in the Voigt profile was used. Additionally, the calculated spectrum was shifted as a whole by +10.65~eV for better comparison with the experimental data.

The spectrum consists of two prominent bands called $L_3$ and $L_2$ which correspond to the SOC components of the created core-hole, i.e.\ $3/2$ and $1/2$ total angular momentum, respectively. Transitions to states $253-1000$ and $1001-1260$ of the $[\text{Fe}(\text{H}_2\text{O})_6]^{3+}$ ion correspond to the $L_3$ and $L_2$ peaks, respectively, in the core PES shown in Fig.~\ref{FePEScore}. The computed core PES with numerical integration according to Eq.~\eqref{PES1} (full calculation) is in rather good agreement with the experiment. Our method reproduces well the $L_2/L_3$ energy splitting, intensity ratio and the general asymmetric shape of the bands with the long tails at the low energy sides. Almost all of the 1260 transitions have notable intensity, hence the bands and corresponding tails are formed by hundreds of lines, most of them being combination transitions. The small discrepancies, e.g., the small peak at $-722.9$~eV and the minimum at $-720.5$~eV, also originate from a large number of transitions of different origin, what hinders a detailed analysis. 
In principle, they could be ascribed to the lack of dynamic correlation (no RASPT2 correction) and to the presence of different species in solution. 
Note that ionization only from $[\text{Fe}(\text{H}_2\text{O})_6]^{2+}$ was considered in computations.
Further, the averaging over several thermally populated electronic initial states could be necessary. 
Finally, due to the high density of states, the tails of the $L_2$ and $L_3$ peaks are very sensitive to the wings of the lineshape function. 
Therefore, an inclusion of more than two different sets of broadening parameters might be necessary as well.

The wave functions of the final states of $[\text{Fe}(\text{H}_2\text{O})_6]^{3+}$ are much more complex in comparison to those of water. First of all, the core-ionized final states do not have a leading contribution from one configuration and represent a combination of many configurations with comparable weights in Eq.~\eqref{CIwfSD}. To illustrate this issue, the weights of the three most important configuration state functions for the quartet final states of $[\text{Fe}(\text{H}_2\text{O})_6]^{3+}$ are shown in the SI. For other spin manifolds the dependences look similar and are not shown. Second, due to strong SOC for the core-ionized states, a pronounced spin-mixing of states within the $LS$-coupling scheme applied here is observed. More  details of this mixing are given in the SI. This implies that the spin of the final states is not well defined and their wave functions represent linear combinations of the sextet, quartet, and doublet spin-states (Eq.~\eqref{SOCWF}). The complex structure of the wave function is reflected in the DOs (Fig.~\ref{FePEScore}b). Here,  most transitions correspond to DOs, where the real and imaginary parts of the $\alpha$ and $\beta$ spin contributions are of comparable magnitude. Thus a calculation of only real DOs with definite spin would produce erroneous intensities. An exception occurs at the rising flank of $L_3$ (DO (3) in Fig.~\ref{FePEScore}b) which can be ascribed to nearly pure quintet-sextet transitions and thus corresponds to photoelectrons with $\beta$ spin. The $\beta$ electron is easily removed because of  the pairing (exchange) energy.

The DOs for different transitions of both the $L_2$ and $L_3$ bands are combinations of pure atomic 2p orbitals of iron and differ in the absolute and relative magnitude of real and imaginary $\alpha$ and $\beta$ components.  This corresponds to the fact that the electron is removed from the core 2p orbitals and, in contrast to water, the series in Eq.~\eqref{photoEl} converge to zero very quickly for all transitions. One example is shown in Fig.~\ref{H2Ocklm}c for the most intense $L_3$ transition. Since only atomic 2p$_{\rm Fe}$ orbitals are contributing to the DOs, the dipole selection rules hold strictly. 

In  Fig.~\ref{FePEScore} we also show the SA results, where the intensities of the transitions are approximated as the norms of DO and thus the computationally demanding integration in Eq.~\eqref{PES1} is avoided. Apparently, the SA gives a PES which substantially deviates from both the experiment and the full calculation, most notably it shows rather intense tails of $L_3$ and $L_2$ bands. This deviations cannot be eliminated by the fitting of broadenings; note that in Fig.~\ref{FePEScore} the same broadening parameters as for the full calculation are used. This result shows that although this approximation is intensively used in CI \cite{Kivimaki_2008,Lisini_1988} and LFM calculations of core PES \cite{de_groot_core_2008} and in RASSCCF/RASSI calculations of valence PES of heavy elements,~\cite{klooster_calculation_2012} it cannot be considered as being generally accurate. This holds especially if the kinetic energy of the ejected electron is relatively low. Thus to obtain accurate L-edge core PES the integration of the DO with the free electron wave function is required. Commonly, the SA is used when the photoelectron momentum angular distribution is not of interest. Our results point to the failure of the SA even in such cases.
%
\section{Conclusions}\label{sec:concl}
We have presented a multi-reference approach to core and valence photoelectron spectra of transition metal complexes, taking into account the essential effects due to the multi-configurational character of the wave function and spin-orbit coupling. This method is an extension of the Dyson orbital formalism, previously applied with TDDFT \cite{di_valentin_gas-phase_2014,mignolet_probing_2013} and EOM-CCSD \cite{melania_oana_dyson_2007,oana_cross_2009} techniques, to  RASSCF/RASSI wave functions.  Thereby, an essential point for efficient computation of the  Dyson orbitals is the biorthonormal MO basis transformation.~\cite{malmqvist_calculation_1986} The proposed protocol includes the numerical integration of the matrix elements, resulting in more reliable intensities as compared with the widely used sudden approximation. 

This present approach is complementary to the theoretical X-ray spectroscopic techniques, which have been recently reported for absorption, fluorescence, and inelastic scattering spectra.~\cite{josefsson_ab_2012,wernet_dissecting_2012,suljoti_direct_2013,bokarev_state-dependent_2013,atak_nature_2013,engel_chemical_2014,pinjari_restricted_2014,Wernet_2015} Having the method describing photon-out and electron-out events on the same footing provides additional tools to address the electronic structure of transition metal complexes.

The computational protocol has been demonstrated for two examples, i.e.\ the valence PES of gaseous water and L-edge core PES of the $[\text{Fe}(\text{H}_2\text{O})_6]^{2+}$ cluster as a model for the $\text{Fe}^{2+}$ ion in aqueous solution. In both cases the agreement between theory and experiment was rather good. In particular for the aqueous ion the RASSCF/RASSI wave function has a rather complex structure. This is reflected in the DO, i.e.\ due to spin-mixing the Dyson orbitals for core-ionization of $[\text{Fe}(\text{H}_2\text{O})_6]^{2+}$ contain complex-valued $\alpha$ and $\beta$ contributions. This immediately implies that the usage of real DOs with definite spin would give erroneous spectra. 

We have contrasted our results with those of the widely used sudden approximation. The latter shows for relatively low excitation energies notable deviations from experimental spectra for the aqueous $\text{Fe}^{2+}$ ion. Hence, in general the numerical integration of the Dyson orbitals with the free electron wave function should not be dismissed.
\begin{acknowledgments}
We acknowledge financial support by the Deanship of Scientific Research (DSR), King Abdulaziz University, Jeddah, (grant No.\ D-003-435) and the Deutsche Forschungsgemeinschaft (grant No.\ KU952/10-1). 
\end{acknowledgments}


\end{document}